\documentclass{cimento}

\usepackage{graphicx}
\usepackage{physics}
\usepackage{bbm}
\DeclareMathOperator{\hc}{H.c.}

\beforemaketitle{\vspace*{-32pt}}
\title{Non-Markovian dynamics of a qubit on the zigzag edge of photonic graphene}

\author{Enrico Di Benedetto\from{}\thanks{Corresponding email address: enrico.dibenedetto@unipa.it}}
\instlist{\inst{} Universit\`{a} degli Studi di Palermo, Dipartimento di Fisica e Chimica -- Emilio Segr\`{e}, via Archirafi 36, I-90123 Palermo, Italy}

\begin{document}

\maketitle

\begin{abstract}
We present an analytical study of the dynamics of a qubit coupled to the zigzag edge of a photonic honeycomb lattice. Leveraging a resolvent operator approach and exact lattice Green's functions, we map the spectrum of qubit-photon bound states and complex poles to uncover the long-time scaling laws of the qubit population. Unlike a qubit coupled to the bulk of graphene, the presence of a zero-energy flat band connecting the Dirac points suppresses long-time branch-cut-induced tails, instead driving damped Rabi oscillations. Potential experimental platforms are briefly discussed.
\end{abstract}
\vspace{-24pt}

\section{Introduction}

Integrating quantum emitters into 2D photonic baths modeled after pristine graphene \cite{castro_neto_electronic_2009} enables the exploration of exotic quantum dynamics, including non-exponential decay and long-range interactions near the Dirac points \cite{gonzalez-tudela_exotic_2018,navarro-baron_photon-mediated_2021}. While bulk implementations in circuit QED \cite{jouanny_high_2025, yang_circuit_2016} or cold atoms \cite{bouscal_systematic_2024, jamadi_direct_2020,milicevic_type_2019} remain challenging, coupling qubits to the edge of these baths is often more feasible. This raises a critical question: how do edge and bulk dynamics differ? 
In this work, we investigate a single quantum emitter coupled to a zigzag graphene edge \cite{tan_edge_2021}. Using a non-perturbative method \cite{lambropoulos_fundamental_2000}, we demonstrate that an edge-localized flat band \cite{benedetto_dipole-dipole_2025, benedetto_emergent_2025} significantly alters the dynamics, triggering damped Rabi oscillations near the Dirac points and removing self-energy branch-cut contributions in the middle of the band.

\begin{figure}[t]
    \centering
    \includegraphics[width=\linewidth]{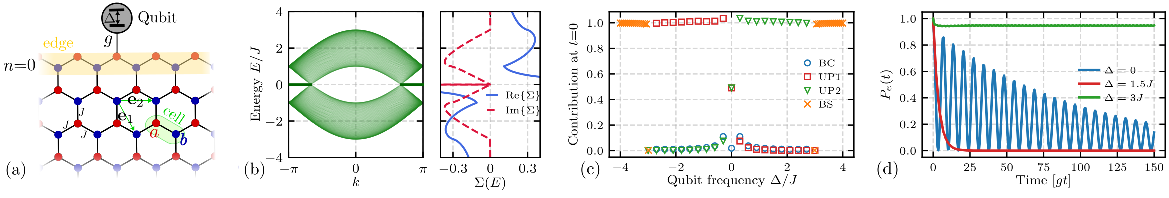}
    \caption{(a) Qubit coupled to the edge located at $n{=}0$ at a cavity in the $A$ sublattice. (b) Left panel: energy spectrum of $H_B$ as a function of $k$. Right panel: real and imaginary part of the self-energy $\mathbf{\Sigma}_e(E{+}i0^+)$. Both panels share the same energy axis. (c) Relative weight of each contribution to the dynamics at $t=0$ at different values of $\Delta/J$ ($g{=}0.2J$). (d) Qubit population over time for $g=0.2J$. We can distinguish three regimes: when the two unstable poles' contributions are relevant (blue curve, $\Delta{=}0$), when a single unstable pole is relevant (red curve, $\Delta{=}1.5J$) and when a bound state appears (green curve, $\Delta{=}3J$).}
    \label{fig:setup}\vspace*{-12pt}
\end{figure}

\section{Model and Hamiltonian}

We consider a 2D honeycomb lattice bath with a zigzag edge [Fig.\ref{fig:setup}(a)], chosen for its peculiar spectral properties \cite{tan_edge_2021}. Modeled via the tight-binding approximation, photons hop between nearest-neighbor cavities at rate $J$. The lattice is generated by primitive vectors $\vb{e}_1 {=} \sqrt{3}(\sqrt{3}/2,1/2)$, $\vb{e}_2 {=} \sqrt{3}(0,1)$ and basis $\vb{d} {=} (1/2,\sqrt{3}/2)$. Site positions are $\vb{R}_{nm\nu} {=} n\vb{e}_1 {+} m\vb{e}_2 {+} \nu \vb{d}$, where $\nu\in\{0,1\}$ denotes the $A$ or $B$ sublattice. The zigzag edge is set at $n{=}0$, imposing open boundaries perpendicular to the edge ($n{\in}\mathbbm{N}$) and periodic boundaries parallel to it ($m{\in}\mathbbm{Z}$). The bath Hamiltonian reads
\begin{equation}
    \label{eq:bath_hamiltonian}
    H_B = J \sum_{\vb{R}} \qty[\qty(a_{\vb{R}+\vb{e}_1} + a_{\vb{R}} + a_{\vb{R}+\vb{e}_2})b_{\vb{R}}^\dag+\hc]\,, 
\end{equation}
where $a_{\vb{R}}$ ($b_{\vb{R}}$) are bosonic annihilation operators for the $A$ ($B$) sublattice. 
A quantum emitter with bare frequency $\Delta$ and annihilation operator $\sigma^-$ is coupled (strength $g$) to an edge cavity $a_{00}$. Under the rotating-wave approximation, the full Hamiltonian of the composite qubit-bath system reads $H = \Delta \sigma^+\sigma^- + H_B + g \qty(\sigma^+a_{00}+\hc)$.

\subsection{Spectrum and normal modes of $H_B$}
Exact diagonalization \cite{benedetto_emergent_2025} yields bulk modes at energies $\omega_{\pm}(k,q) = \pm \sqrt{J^2+\mathcal{J}^2(k) + 2 J\mathcal{J}(k)\cos{q}}=\pm J\,\omega(k,q)$ for $0 {<} q {<} \pi$, with $\mathcal{J}(k) = 2J\cos(k/2)$. These modes form a continuous band in $\comm{-3J}{3J}$ with two zero-energy Dirac cones at $(k,q){=} (\pm2\pi{/}3,\pi)$ [Fig.\ref{fig:setup}(b)]. 
For $2\pi{/}3 {<} \abs{k} {\leq} \pi$, the system possesses zero-energy topological edge modes. These modes are exponentially localized along the open direction (with a localization length $\lambda_k^{-1} {=} \ln|J/\mathcal{J}(k)|$) and form a zero-energy flat band ($\omega_{\rm FB}{=}0$) strictly on the $A$ sublattice.

\subsection{Self-energy} 
The exact qubit dynamics are governed by the self-energy $\mathbf{\Sigma}(z){=}\\ g^2 \mel{\rm vac}{a_{00}(z-H_B)^{-1}a_{00}^\dagger}{\rm vac}$, for which we derive a closed-form expression
\begin{equation}
    \label{self-energy}
    \mathbf{\Sigma}(z) = \frac{g^2}{2z} \qty[ \frac{9+4t-2t^2}{6}I_0(t) + \frac{I_{0}(t_0)}{4} - \frac{I_1(t)-I_{1}(t_0)}{2} + \frac{3+2t}{6} ] + \frac{\Omega^2}{z}\,,
\end{equation}
where $t(z)=(z^2-3)/2$, $t_0=-3/2$, and $\Omega=g\sqrt{\sqrt{3}/\pi-1/3}$ is the effective coupling to the flat band. The first term captures the bulk mode contribution, while the second captures the flat band. The functions $I_{0,1}$ \cite{horiguchi_lattice_1972} depend on complete elliptic integrals $\Tilde{K}(z)$ and $\Tilde{E}(z)$ as
\begin{equation}
    \begin{split}
    I_0(z) &= \frac{C}{2\pi}\Tilde{K}(\beta), \\
    I_1(z) &= \frac{C}{2\pi} \qty[ \frac{3(\alpha^2 - 2)(\beta^2 + \alpha^2 - 2) + 2(\beta^2 - 2\alpha^2 + \alpha^4)z}{3(\beta^2 - \alpha^2)\alpha^2} \tilde{K}(\beta) + \frac{4(\alpha^2 - 1) \tilde{E}(\beta)}{(\beta^2 - \alpha^2)\alpha^2} ]\\
    &- \frac{2(\beta^2 - 2\alpha^2 + \alpha^4)}{3(\beta^2 - \alpha^2)\alpha^2},
    \end{split}
\end{equation}
with parameters $\alpha^2(z) = 4(1 {-} \sqrt{2z {+} 3})^{-2}$, $C(z) = 8(\sqrt{2z + 3} {-} 1)^{-3/2} (\sqrt{2z {+} 3} {+} 3)^{-1/2}$, and $\beta(z) = C(z)(2z + 3)^{1/4}/2$. To avoid branch cuts at $z \in \{-3J, -J, 0, J, 3J\}$, $\Tilde{K}$ and $\Tilde{E}$ are analytically continued beyond the first Riemann sheet.

\section{Exact dynamics of the qubit}

For an initially-excited qubit, the population amplitude $c_e(t)$ is computed via the resolvent operator \cite{lambropoulos_fundamental_2000}
\begin{equation}
    c_e(t) = {-}\frac{1}{2\pi i}\int_{-\infty}^\infty \dd E\,G_e(E{+}i0^+)  e^{-iEt}\,,
\end{equation}
where $G_e(z) = \qty[z{-}\Delta{-}\mathbf{\Sigma}(z)]^{-1}$. Closing the contour in the lower half-plane decomposes the dynamics into three distinct contributions
\begin{equation}
    c_e(t) = \sum_{\{\omega_{\rm BS}\}} r_{\rm BS}\, e^{-i\omega_{\rm BS}t} + \sum_{\{\omega_{\rm UP}\}} r_{\rm UP}\, e^{-i\omega_{\rm UP}t} + c_{\rm BC}(t)\,,
\end{equation}
comprising real bound states ($r_{\rm BS}$), unstable complex poles ($r_{\rm UP}$), and branch cut detours ($c_{\rm BC}$). Their relative weights at $t=0$ are plotted in Fig.\ref{fig:setup}(c).

\subsection{Real Bound States} 
Real solutions to $E{-}\Delta{-}\mathbf{\Sigma}(E){=}0$ occur outside the continuous band ($|\omega_{\rm BS}| {>} 3J$). Because $\mathbf{\Sigma}(E)$ is monotonically increasing \cite{shi_bound_2016}, a single bound state emerges only when the qubit is tuned near the upper or lower band edges ($\Delta {>} J[3 {-} 0.297(g/J)^2]$ or $\Delta {<} {-}J[3 {-} 0.297(g/J)^2]$). This overlap leads to \textit{fractional decay}, where the population stabilizes at $\abs{r_{\rm BS}}^2$ [Fig.\ref{fig:setup}(d), green curve].

\subsection{Unstable Poles} 
For $|\text{Re}\{z\}| {\ll} \Omega$, the flat band dominates the self-energy ($\mathbf{\Sigma}(z) \simeq \Omega^2/z$). The pole equation simplifies to $z {-} \Delta {-} \Omega^2/z {=} 0$, yielding $z_{\pm} = (\Delta {\pm} \sqrt{\Delta^2 {+} 4\Omega^2})/2$. The flat band acts as an \textit{effective cavity} coupled to the qubit. Including the bulk contribution shifts these roots into the lower complex half-plane ($\omega_{\rm UP1,2}$), driving damped Rabi oscillations when tuned near resonance. Conversely, for $|\Delta| {\gg} \Omega$, one pole approaches zero, preserving population trapping independent of system size [Fig.\ref{fig:setup}(d), blue and red curves].

\subsection{Branch Cuts} 
While band-edge power-law decays are typically overshadowed by fractional decay \cite{gonzalez-tudela_markovian_2017}, the Dirac point ($E=0$) tail dominates long-time non-exponential behavior. Expanding the analytically continued Green's function near the origin yields the asymptotic behavio:
\begin{equation}
    c_{\rm BC}(t) \simeq \qty(\frac{g}{J})^2 \frac{\sqrt{3}}{\pi} \frac{1}{(\Omega t)^4}\,.
\end{equation}
This $t^{-4}$ decay sharply contrasts with the $1/\log(t)$ scaling characteristic of bulk graphene \cite{gonzalez-tudela_exotic_2018}. However, it remains practically unobservable, masked entirely by Rabi oscillations ($\Delta<\Omega$) or fractional decay ($\Delta>\Omega$) [as confirmed by Fig.\ref{fig:setup}(d)].

\section{Discussion}
The qubit population $P_e(t)$ is driven by the competition between these three spectral components. At short to intermediate times, unstable poles $\omega_{\rm UP1,2}$ drive damped Rabi oscillations with a frequency proportional to the effective vacuum Rabi splitting $\sqrt{\Delta^2 + 4\Omega^2}$. At long times, fractional decay dominates, driven either by a real bound state or a trapping unstable pole. 

This edge-coupled topology is readily accessible in modern cQED platforms using transmon qubits and high-impedance cavity arrays \cite{jouanny_high_2025}, where edge-coupling is far easier to implement than bulk-coupling. The emergence of the flat band fundamentally alters the system's evolution, replacing bulk branch-cut tails with Rabi oscillations. This highlights a powerful method for engineering novel quantum phenomenology simply by modifying lattice boundary conditions.

\acknowledgments
The author thanks F. Ciccarello and F. Roccati for stimulating discussions, and acknowledges financial support from European Union-Next Generation EU through projects: Eurostart 2022, PRIN 2022-PNRR No. P202253RLY, and THENCE-Partenariato Esteso NQSTI-PE00000023-Spoke 2.

\end{document}